\documentclass[12pt]{article}
\usepackage{amsmath,epsfig,amssymb,eufrak}
\usepackage{amscd,latexsym}
\usepackage{color}

\date{}

\def\xlf{\raisebox{+0.2em}{\color{red}\boldmath{$\chi$}}\hspace{-0.2ex}\raisebox{-0.2em}{\color{green}L}
\hspace{-1.5ex}\raisebox{+0.14em}{\color{blue}F}\hspace{2mm}}

\def\lsi{\raise0.3ex\hbox{$<$\kern-0.75em\raise-1.1ex\hbox{$\sim$}}}
\def\gsi{\raise0.3ex\hbox{$>$\kern-0.75em\raise-1.1ex\hbox{$\sim$}}}

\newcommand{\kcpi}{\kappa_c^{\rm pion}}
\newcommand{\kcP}{\kappa_c^{~\!\!\!_{\rm PCAC}}}

\input macros.sty      
\input eqalign.sty     

\begin{document}

\begin{titlepage}

\title{
  {\vspace{-0cm} \normalsize
  \hfill \parbox{40mm}{DESY 05-113\\
                       SFB/CPP-05-29\\July 2005}}\\[10mm]
Quenched Scaling of Wilson twisted mass fermions}  
\author{ K.~Jansen$^{\, 1}$, M.~Papinutto$^{\, 1}$, A.~Shindler$^{\, 1}$, \\
C.~Urbach$^{\, 1,2}$ and I.~Wetzorke$^{\, 1}$\\ 
\\
   {\bf \xlf Collaboration}\\
\\
{\small $^{1}$  John von Neumann-Institut f\"ur Computing NIC,} \\
{\small         Platanenallee 6, D-15738 Zeuthen, Germany} \\ \ \\
{\small $^{2}$  Institut f\"{u}r Theoretische Physik, Freie Universit\"{a}t Berlin,} \\
{\small Arnimallee 14, D-14195 Berlin, Germany}
}

\maketitle

\begin{abstract}
We investigate the scaling behaviour of quenched Wilson twisted mass fermions
at maximal twist applying two definitions of the critical mass. 
The first definition
uses the vanishing of the pseudoscalar meson mass $m_{\rm PS}$ while the 
second employs the vanishing of the PCAC quark mass $m_{\rm PCAC}$. 
We confirm in both cases the expected $O(a)$ improvement. 
In addition, we show that the PCAC quark
mass definition leads to substantially reduced 
$O(a^2)$ cut-off effects 
even when the pseudoscalar meson mass $m_{\rm PS}$ is as small as 
$270$ MeV. At a fixed value of $m_{\rm PS}$ we perform continuum limits for the
vector meson mass $m_V$ and for the pseudoscalar decay constant 
$f_{\rm PS}$ and discuss the renormalisation constant $Z_V$ of the vector 
current.  
\vspace{0.75cm}
\noindent
\end{abstract}

\end{titlepage}

\section{Introduction}

Knowing the scaling behaviour of a given lattice action is an important step
towards performing a controlled continuum limit of physical observables. 
The important result that Wilson fermions with a twisted mass 
term added \cite{Frezzotti:2000nk} lead, at full twist, to an automatic 
$O(a)$-improvement of correlators \cite{Frezzotti:2003ni} without the need of
additional improvement terms makes it particularly interesting to test this 
formulation of lattice QCD with numerical simulations. 
In this paper we will use lattice spacings ranging from 
$a=0.05$ fm to $a=0.17$ fm and pseudoscalar meson masses from
$m_{\rm PS}=1180$ MeV down to
$m_{\rm PS}=270$ MeV in order to perform a detailed scaling test of Wilson 
twisted mass fermions. In this way, we extend a first scaling test
\cite{Jansen:2003ir} substantially. 
All our simulations presented here are done in the quenched approximation,
see also 
\cite{Bietenholz:2004wv,Jansen:2005gf,Abdel-Rehim:2004gx,Abdel-Rehim:2005gz} 
for further quenched results of Wilson twisted mass QCD. 
Let us mention that also full QCD simulations with this 
approach have already been performed and proved to be very useful 
in studying the phase structure of lattice QCD with Wilson fermions
\cite{Farchioni:2004us,Farchioni:2004ma,Farchioni:2004fs,Ilgenfritz:2003gw,Sternbeck:2003gy}. 

A basic ingredient to have automatic $O(a)$-improvement is the tuning of
the bare quark mass to its critical value. In the language of
twisted mass QCD (tmQCD) this procedure corresponds to fix the twist 
angle to $\pi/2$. 
In the present work, we will employ two 
definitions for the critical mass. The first is the point where 
the pseudoscalar meson mass vanishes, the second, where the 
PCAC quark mass vanishes. In the following we will refer to the first
situation as the ``pion definition'' and to the second situation 
as the ``PCAC definition'' of the critical point.
Both definitions should lead to $O(a)$-improvement, but 
they can induce very different $O(a^2)$ effects, in particular at 
small pseudoscalar meson masses. Indeed, 
in ref.~\cite{Bietenholz:2004wv,Bietenholz:2004sa} we reported that 
the pion definition can have substantial $O(a^2)$ effects which are
amplified when the quark mass becomes small and violates the 
inequality \mbox{$\mu > a\Lambda^2$} (where $\mu$ is the bare quark mass,
i.e. the parameter which provides at full twist a mass to the 
pseudoscalar meson). 
On the other hand, when the PCAC definition of the critical
mass is used, these particular kind of $O(a^2)$ cut-off effects are 
dramatically reduced as was demonstrated in \cite{Jansen:2005gf,Abdel-Rehim:2005gz}. 
The effect of reducing $O(a^2)$ artefacts has been theoretically 
studied in chiral perturbation theory in 
refs.~\cite{Aoki:2004ta,Sharpe:2004ny} and put on a more general basis 
in ref.~\cite{Frezzotti:2005gi}.

In this paper, we employ the pion and the PCAC definitions of the 
critical mass to study the scaling behaviour of the pseudoscalar meson 
decay constant and the vector meson mass. The renormalisation 
constant $Z_V$ of the vector current is also presented for three
values of the lattice spacing. 

\section{Wilson twisted mass fermions}
Wilson twisted mass fermions can be 
arranged to be $O(a)$ improved without employing specific improvement terms
\cite{Frezzotti:2003ni}. The Wilson tmQCD action in the twisted basis 
can be written as
\begin{equation}
  \label{tmaction}
  S[U,\psi,\bar\psi] = a^4 \sum_x \bar\psi(x) ( D_W + m_0 + i \mu
\gamma_5\tau_3 ) \psi(x)\; ,
\end{equation}
where the Wilson-Dirac operator $D_{\rm W}$ is given by
\be
D_{\rm W} = \sum_{\mu=0}^3 \frac{1}{2} 
[ \gamma_\mu(\nabla_\mu^* + \nabla_\mu) - a \nabla_\mu^*\nabla_\mu]
\label{Dw}
\ee
and $\nabla_\mu$ and $\nabla_\mu^*$ denote the usual forward
and backward derivatives. We refer to \cite{Bietenholz:2004wv} for further
unexplained notations. The definition of the critical mass $m_c$ will be
discussed in detail in the next section.

We extract pseudoscalar and vector meson masses from the correlation
functions at full twist ($m_0=m_c$):
\begin{eqnarray}
C_P^a(x_0) &=& a^3\sum_{\mathbf x} \langle P^a(x)P^a(0)\rangle \quad a=1,2\label{cp}\\
C_A^a(x_0) &=& \frac{a^3}{3}\sum_{k=1}^3\sum_{\mathbf x} \langle A_k^a(x)A_k^a(0)\rangle 
\quad a=1,2 \label{ca}\\
C_T^a(x_0) &=& \frac{a^3}{3}\sum_{k=1}^3\sum_{\mathbf x} \langle T_k^a(x)T_k^a(0)\rangle
\quad a=1,2 \label{ct}
\end{eqnarray}
where we consider the usual local bilinears 
$P^a=\bar{\psi}\gamma_5\frac{\tau^a}{2}\psi$,
$A_k^a=\bar{\psi}\gamma_k\gamma_5\frac{\tau^a}{2}\psi$ and
$T_k^a=\bar{\psi}\sigma_{0 k}\frac{\tau^a}{2}\psi$.

The untwisted PCAC quark mass $m_{\textrm{PCAC}}$ can be extracted from the ratio
\be
m_{\textrm{PCAC}}=\frac{\sum_{\mathbf x}\langle
\partial_0 A_0^a(x)\; P^a(0)\rangle}{2\sum_{\mathbf x}\langle 
P^a(x)P^a(0)\rangle}\quad a=1,2\; .
\label{mPCAC}
\ee
Using the exact lattice PCVC relation
\be
\langle\partial^*_\mu \tilde V^a_\mu(x) O(0)\rangle= -2 \mu \epsilon^{3ab}
\langle P^b(x) O(0)\rangle \qquad a=1,2
\label{PCVC}
\ee
(where $\partial^*_\mu$ is the lattice backward derivative, $\tilde
V^a_\mu$ is the point-splitted vector current and $O$ is a local lattice
operator) we can also compute the
pseudoscalar meson decay constant at maximal twist without requiring 
any renormalisation constant 
(see \cite{Frezzotti:2001du,DellaMorte:2001tu,Jansen:2003ir}):
\be
f_{\rm PS}=\frac{2\mu}{m_{\rm PS}^2} | \langle
0 | P^a | {PS}\rangle | \qquad a=1,2 \; ,
\label{indirect}
\ee
where $m_{\rm PS}$ is the charged pseudoscalar mass and $| {PS}\rangle$ denotes
the corresponding pseudoscalar state. 

If one now is interested in using the local definition of the vector current 
$V^a_\mu=\bar{\psi}\gamma_\mu\frac{\tau^a}{2}\psi$,  
$\tilde V^a_\mu$ has to be substituted with $Z_V V^a_\mu$ in 
eq.~(\ref{PCVC}) and the relation
is now valid only up to order $O(a^2)$ lattice artifacts.
This relation allows for a determination of $Z_V$ through 
\be
Z_V=\lim_{\mu\rightarrow 0} \frac{-2 \mu \epsilon^{3ab} \sum_{\mathbf x}\langle P^b(x) 
P^b(0)\rangle} {\sum_{\mathbf x}\langle \tilde\partial_\mu V^a_\mu(x) P^b(0)\rangle} \qquad a=1,2 \,,
\label{eq:Z_V}
\ee
where $\tilde\partial_\mu$ is the symmetric lattice derivative.

\section{Definition of the critical mass}

The Wilson tmQCD action of eq.~(\ref{tmaction}) can be studied in the full
parameter space $(m_0,\mu)$. A special case arises, however, when
$m_0$ is tuned to its critical value $m_{\mathrm{c}}$.
In such, and only in such a situation, all physical quantities are, or
can easily be, $O(a)$ improved. The critical mass, or 
alternatively the critical hopping parameter 
$\kappa_c = (2 a m_c+8)^{-1}$, has thus to be fixed in the 
actual simulation to achieve a twist angle of $\pi/2$.

One possible definition of $\kappa_c$ is to perform at vanishing twisted mass
parameter an extrapolation of $(m_{\rm PS} a)^2 \to 0$.  
However, as we said in the introduction, it leads to the presence of 
large lattice artifacts which are amplified at small quark masses. 
A better definition of $\kappa_c$ (which reduces this kind of $O(a^2)$
artifacts) can be obtained by first determining at fixed non-zero 
twisted mass parameter the value of $\kappa_c(\mu a)$, 
where the PCAC quark mass of eq.~(\ref{mPCAC}) vanishes.  
The non-zero
value of the twisted mass allows a safe interpolation in this case.
As a further step, an only short linear extrapolation of 
$\kappa_c(\mu a)$ from small values of $\mu a$ to $\mu a = 0$ yields 
a definition of $\kappa_c$ which is expected 
\cite{Aoki:2004ta,Sharpe:2004ny,Frezzotti:2005gi} 
to lead to $O(a^2)$ lattice artefacts which remain small even at small masses.
In ref.~\cite{Jansen:2005gf} we gave an 
example of such a computation of the critical mass and found 
that indeed the cut-off effects at small quark masses are 
substantially reduced when compared to results obtained from the pion 
definition of $\kappa_c$.

Recently, a definition of maximal twist from parity conservation has also been 
investigated in \cite{Abdel-Rehim:2005gz}. There, the critical masses 
$m_c(\mu a)$ 
were not extrapolated to $\mu a = 0$, but were used at the respective
twisted mass parameter at which they were determined.

\section{Numerical results}
In this section we will provide a comparison of Wilson twisted mass
results for the pseudoscalar meson mass and the pseudoscalar meson 
decay constant as obtained using 
the pion and the PCAC
definitions of $\kappa_c$. Moreover we will present results for the 
vector meson mass and the renormalisation constant $Z_V$ 
as obtained using only the latter definition of $\kappa_c$. 
Our simulations are performed for
6 different values of the lattice spacing, $ 0.05\le a\le 0.17$ fm.
The fermion matrix was inverted for a number of bare quark masses 
in a corresponding pseudoscalar meson mass 
range of $270 \mathrm{~MeV} < m_{\rm PS} < 1180 \mathrm{~MeV}$
using a multiple mass solver (MMS), which is explained in the appendix,
on $O(200)-O(600)$ (depending on the $\beta$) gauge field configurations
generated with the Wilson plaquette gauge action. 
See table~\ref{table:data} for an overview of the simulation points and
table~\ref{table:kappa} for the values of $\kappa_c$ obtained with
both determinations.

\begin{table}[!t]
\begin{center}
\begin{tabular}{|c||c|c|c|c|c|c|}
\hline
\hline
$\beta$ & 5.70 &  5.85 & 6.00 & 6.10 & 6.20  & 6.45 \\
\hline   
$a$ (fm) & 0.171 & 0.123 & 0.093 & 0.079 & 0.068 & 0.048 \\
$r_0/a$ & 2.930 & 4.067 & 5.368 & 6.324 & 7.360 & 10.458\\
$L/a$ & 12 & 16 & 16 & 20 & 24 & 32\\
$T/a$ & 32 & 32 & 32 & 40 & 48 & 64\\
\hline
\hline
&\multicolumn{6}{|c|}{pion definition ($\kcpi$)}\\
\hline
$N_{\rm meas}$ & 600 & 378 & 387 & 300 & 260 & 182 \\
\hline
$\mu_1 a$& 0.0070 & 0.0050 & 0.0038 & 0.0032 & 0.0028 & 0.0020\\
$\mu_2 a$& 0.0139 & 0.0100 & 0.0076 & 0.0064 & 0.0055 & 0.0039\\
$\mu_3 a$& 0.0278 & 0.0200 & 0.0151 & 0.0128 & 0.0111 &       \\
$\mu_4 a$& 0.0556 & 0.0400 & 0.0302 & 0.0257 & 0.0221 &       \\
$\mu_5 a$& 0.0834 & 0.0600 & 0.0454 & 0.0385 & 0.0332 &       \\
$\mu_6 a$& 0.1112 & 0.0800 & 0.0605 & 0.0514 & 0.0442 &       \\
$\mu_7 a$& 0.1390 & 0.1000 & 0.0756 & 0.0642 & 0.0553 &       \\
\hline                                                     
\hline                                                     
&\multicolumn{6}{|c|}{PCAC definition ($\kcP$)}\\
\hline
$N_{\rm meas}$ & 600 & 500 & 400 & & 300 & \\
\hline
$\mu_1 a$& 0.0070 & 0.0050 & 0.0038 &   & 0.0028 &\\
$\mu_2 a$& 0.0139 & 0.0100 & 0.0076 &   & 0.0055 &\\
$\mu_3 a$& 0.0278 & 0.0200 & 0.0151 &   & 0.0111 &\\
$\mu_4 a$& 0.0556 & 0.0400 & 0.0302 &   & 0.0221 &\\
$\mu_5 a$& 0.0834 & 0.0600 & 0.0454 &   & 0.0332 &\\
$\mu_6 a$& 0.1112 & 0.0800 & 0.0605 &   & 0.0442 &\\
$\mu_7 a$& 0.1390 & 0.1000 & 0.0756 &   & 0.0553 &\\\hline
$\mu_8 a$& 0.0200 & 0.0144 & 0.0109 &   & 0.0080 &\\
$\mu_9 a$& 0.0420 & 0.0302 & 0.0228 &   & 0.0166 &\\\hline
\hline
\end{tabular}
\end{center}
\caption{\it Simulation parameters and statistics ($N_{\rm meas}$)}
\label{table:data}
\end{table}

\begin{table}[!t]
\begin{center}
\begin{tabular}{|c||c|c|}
\hline
\hline
$\beta$ & $\kcpi$  &  $\kcP$ \\
\hline
\hline
5.7  & 0.169198(48) &  0.171013(160) \\
5.85 & 0.161662(17) &  0.162379(93)  \\ 
6.0  & 0.156911(35) &  0.157409(72)  \\
6.1  & 0.154876(10) &       --         \\
6.2  & 0.153199(16) &  0.153447(32)  \\
6.45 & 0.150009(11) &       --         \\
\hline
\hline
\end{tabular}
\end{center}
\caption{\it Critical values of the hopping parameters obtained from the
vanishing of the pseudoscalar meson mass ($\kcpi$) and from the
vanishing of the PCAC mass ($\kcP$).}
\label{table:kappa}
\end{table}

In table~\ref{table:mpi} we present the values of the pseudoscalar
masses (in lattice units) for a set of 7 values of
$\mu_i$ $(i=1,\ldots,7)$ of the twisted mass, 
chosen at the different $\beta$'s in order to roughly give 
the same value of the pseudoscalar mass in physical units.
For the PCAC definition of $\kappa_c$ we added two more intermediate
masses in the MMS, denoted as $\mu_8$ and $\mu_9$ in the tables.
In table~\ref{table:fpi} we present the corresponding values for $f_{\rm PS}$,
as obtained from eq.~(\ref{indirect}). The goal of this paper is to
perform the continuum extrapolation of $f_{\rm PS}$ at a fixed value
of $m_{\rm PS}$, for a number of values of  $m_{\rm PS}$. The first step of this
procedure consists in interpolating the value of $f_{\rm PS} r_0$ for the
chosen values of $m_{\rm PS} r_0$. These values are close to the simulated
ones, and so even a linear interpolation is usually sufficient. Since we do
not want to extrapolate out of the range of simulated masses, the lowest mass
that can be reached is given by the highest value of $m_{\rm PS} r_0$ (in both
sets of data) corresponding to $\mu_1$, which is given by the point at
$\beta=6.45$ with the pion definition of $\kappa_c$. This point corresponds to
$m_{\rm PS}=297$ MeV
\footnote{Throughout this work we use the value $r_0=0.5$ fm.}. 

\begin{figure}[htb]
\vspace{-0.0cm}
\begin{center}
\epsfig{file=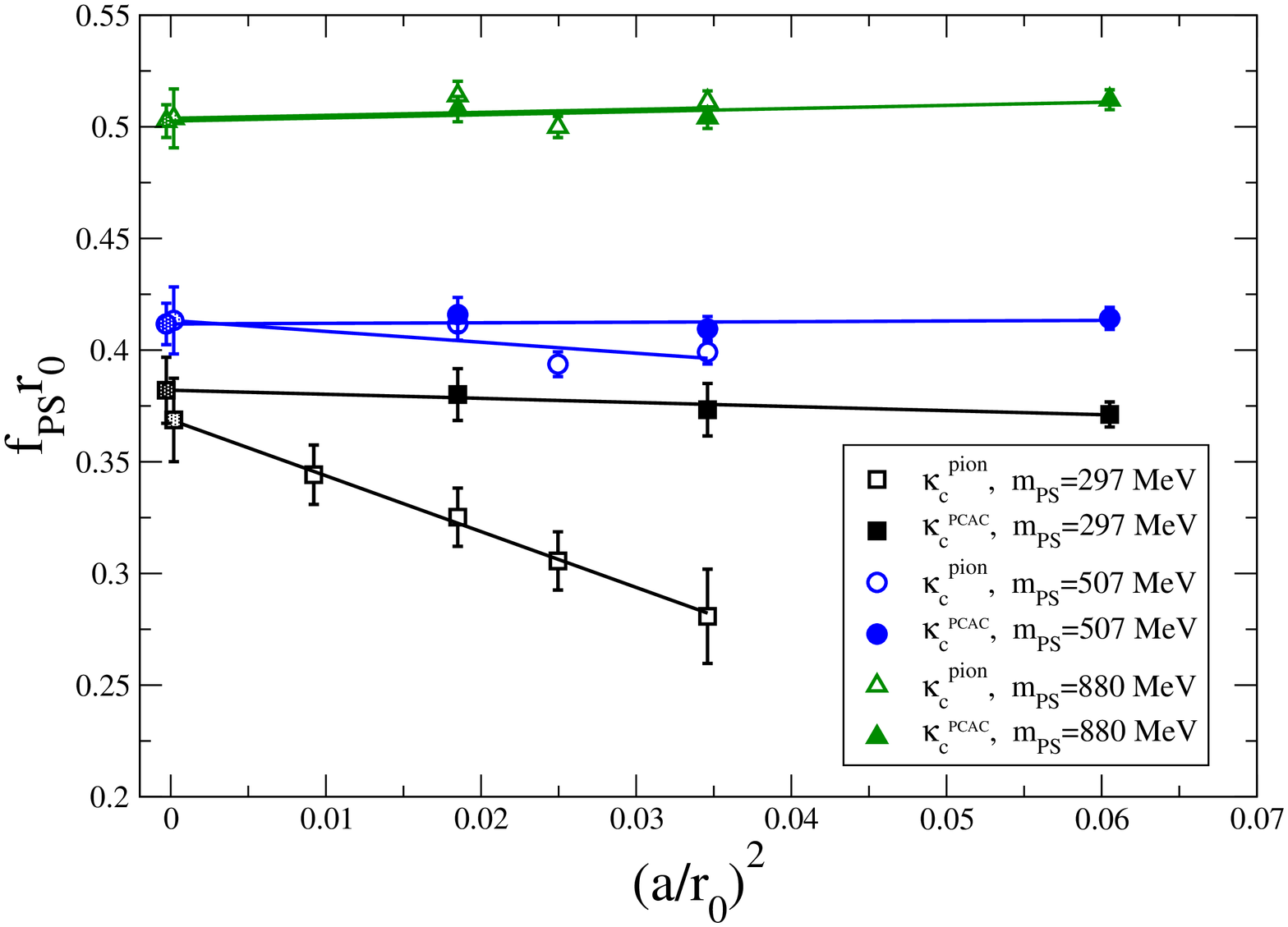,angle=270,width=0.8\linewidth}
\epsfig{file=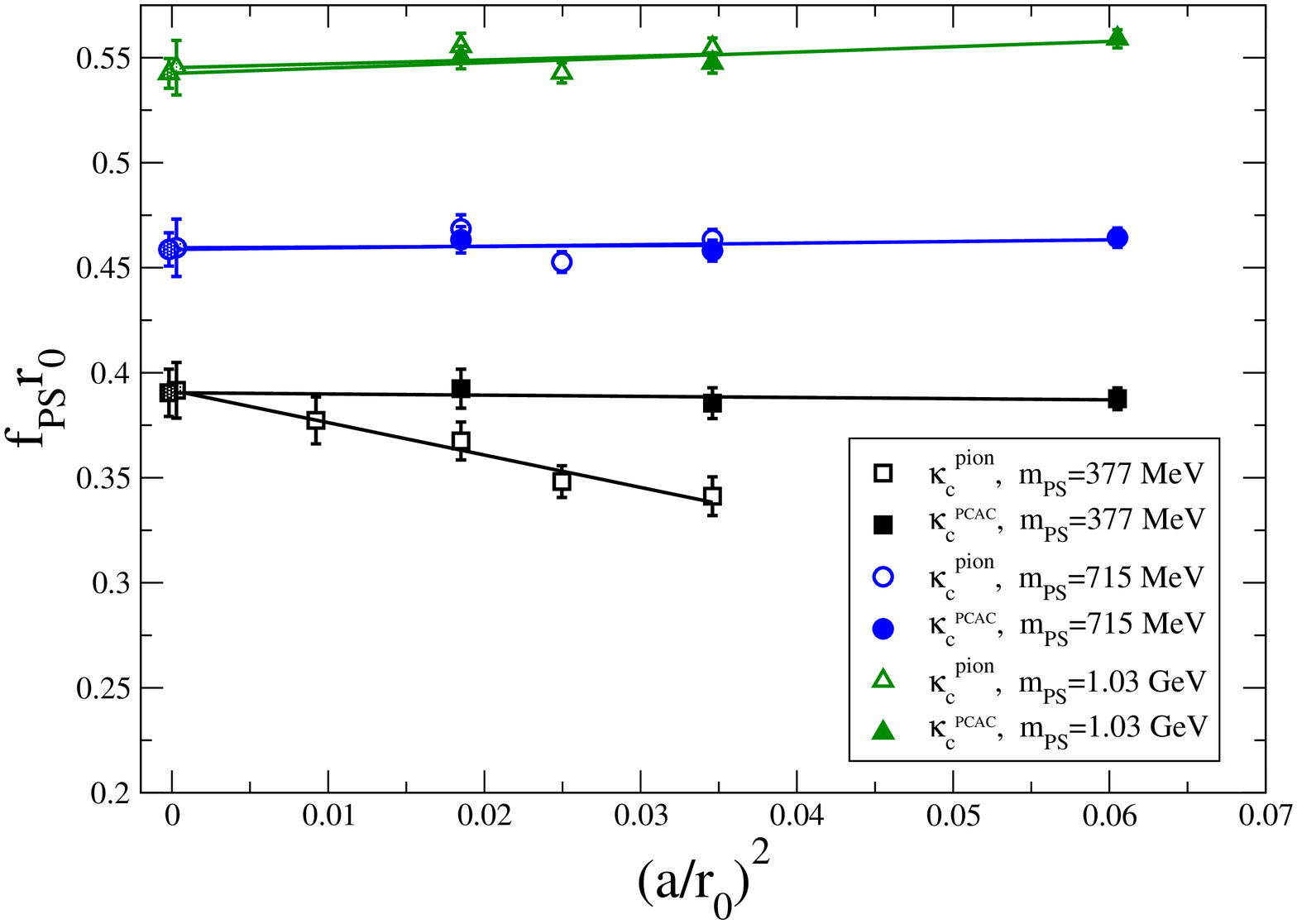,angle=270,width=0.8\linewidth}
\end{center}
\vspace{-0.0cm}
\caption{$r_0f_{\rm PS}$ as a function of $(a/r_0)^2$ using the 
pion definition (open symbols) and the PCAC definition (filled symbols)
of the critical mass; fits are performed with a linear function in
$(a/r_0)^2$ separately for each set.
\label{fig:fpi}}
\end{figure}

In fig.\ref{fig:fpi}  
we plot $f_{\rm PS} r_0$ as a function of $(a/r_0)^2$ for pseudoscalar masses
in the range 297-1032 MeV, obtained with both definitions of $\kappa_c$.
For values of beta large enough, the values $f_{\rm PS} r_0$ show, with both 
definitions of $\kappa_c$, a linear behaviour in $(a/r_0)^2$. This nicely
demonstrates the $O(a)$ improvement for both definitions of $\kappa_c$. 
However, for the pion definition we notice that the effects in 
$O(a^2)$ are rather large, in particular at small pseudoscalar meson masses of 
$297$ MeV and $377$ MeV.
In contrast, the PCAC definition reveals an almost flat behaviour as a function
of $(a/r_0)^2$  even at these small pseudoscalar meson  masses.  
This confirms the results of~\cite{Jansen:2005gf} that the PCAC definition
provides  a better definition of the critical mass with substantially improved
scaling properties of physical observables, especially at small quark 
masses.
In order to take the continuum limit we identify the scaling
region where the data are well described by corrections linear in
$(a/r_0)^2$.
This turns out to start at $\beta=6.0$ with the pion definition of $\kappa_c$ 
and at $\beta=5.85$ with the PCAC definition of $\kappa_c$. The values in the
continuum limit are obtained separately by performing linear fits to the data
in these two regions. The results of these fits are shown in fig.~\ref{fig:fpi}
together with the data. It is very reassuring that these independent linear
fits lead to completely consistent continuum values.

\begin{figure}[htb]
\vspace{-0.0cm}
\begin{center}
\epsfig{file=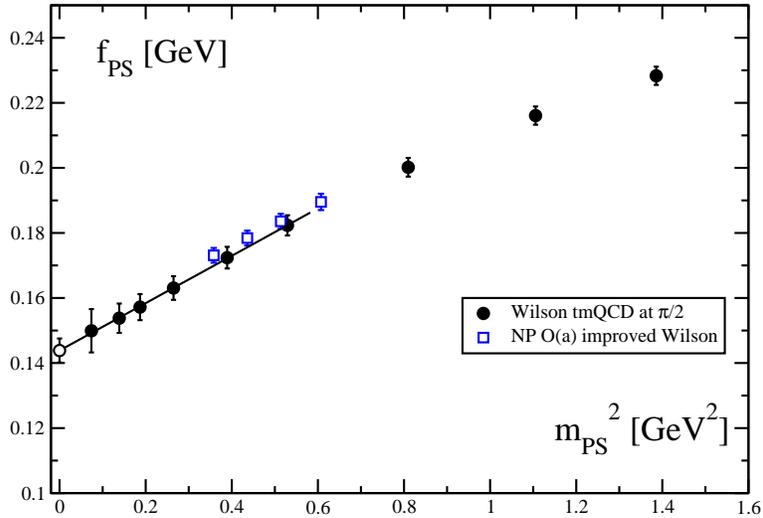,angle=270,width=0.8\linewidth}
\end{center}
\vspace{-0.0cm}
\caption{Continuum behaviour of $f_{\rm PS}$ (filled circles) as a function 
of the pseudoscalar mass squared (in physical units). We also
plot the continuum limit obtained with non-perturbatively improved
Wilson fermions (open squares)~\cite{Garden:1999fg}.
\label{fig:fpscont}}
\end{figure}

In view of the better scaling behaviour of the data obtained with the PCAC
definition of $\kappa_c$, for which the continuum extrapolation 
is safely under control already with data corresponding to values of 
$\beta$ in the range
$[5.85-6.2]$, we decided to use only these data to obtain the final
results in the continuum limit. As already said, data with the PCAC definition
of $\kappa_c$ are affected by much smaller lattice artifacts at small
masses. In particular the minimal pseudoscalar meson mass that corresponds to
$\mu_1$ is now 272 MeV. We thus choose nine pseudoscalar meson masses in the
range  272-1177 MeV and extrapolate the corresponding values of $f_{\rm PS}$ to
the continuum limit. The results are presented in table~\ref{table:cont} 
and fig.~\ref{fig:fpscont}.    
The values of $f_{\rm PS}$ show a linear behaviour down to pseudoscalar 
meson masses of 
about $270$ MeV without signs of chiral logarithms which, for this 
quantity, should only appear in quenched chiral perturbation theory
(q$\chi$PT) beyond one loop. In the same figure we also plot the
continuum values as obtained from results of the ALPHA Collaboration 
with non-perturbatively improved Wilson fermions~\cite{Garden:1999fg}. Notice
that for this action, due to the presence of exceptional configurations, 
the smallest pseudoscalar meson mass that could be simulated was
above 550 MeV.  
In table~\ref{table:cont} we also present 
the results of a linear extrapolation of our data to the chiral limit 
(performed on the six
smallest masses) through which we compute the values of the pion and
kaon decay constant $f_\pi$ and $f_K$ (the latter in the $SU(3)$
symmetric limit). The ratio of the two gives $f_K/f_\pi=1.11(4)$, which
is $10\%$ smaller than what is obtained experimentally. This is however
consistent with what was observed in previous quenched 
calculations~\cite{Heitger:2000ay}.    

\begin{table}[!t]
\begin{center}
\begin{tabular}{|c||c|c|c|c|c|c|}
\hline
\hline
$\beta$ & 5.70 &  5.85 & 6.00 & 6.10 & 6.20  & 6.45 \\   
\hline
\hline
&\multicolumn{6}{|c|}{$m_{\rm PS} a$ ($\kcpi$)}\\
\hline
$\mu_1 a$& 0.2455(23)&  0.1682(26)&  0.1385(66)&   0.1129(41)&   0.1004(27) &0.0720(28)\\
$\mu_2 a$& 0.3237(16)&  0.2256(22)&  0.1764(42)&   0.1482(27)&   0.1298(23) &0.0914(27)\\
$\mu_3 a$& 0.4434(11)&  0.3122(19)&  0.2373(32)&   0.2030(21)&   0.1768(17) &\\
$\mu_4 a$& 0.6272(09)&  0.4452(14)&  0.3335(22)&   0.2865(15)&   0.2463(15) &\\
$\mu_5 a$& 0.7767(09)&  0.5535(12)&  0.4134(17)&   0.3534(13)&   0.3037(13) &\\
$\mu_6 a$& 0.9074(09)&  0.6488(13)&  0.4839(16)&   0.4130(13)&   0.3546(12) &\\
$\mu_7 a$& 1.0255(08)&  0.7358(12)&  0.5491(14)&   0.4676(12)&   0.4021(11) &\\
\hline                                                     
\hline                                                     
&\multicolumn{6}{|c|}{$m_{\rm PS} a$ ($\kcP$)}\\
\hline
$\mu_1 a$& 0.2323(18) & 0.1640(23) &  0.1217(66) &   &  0.0934(24)&\\
$\mu_2 a$& 0.3245(15) & 0.2289(17) &  0.1708(50) &   &  0.1276(21)&\\
$\mu_3 a$& 0.4598(12) & 0.3232(13) &  0.2396(33) &   &  0.1779(18)&\\
$\mu_4 a$& 0.6564(10) & 0.4606(11) &  0.3403(22) &   &  0.2492(13)&\\
$\mu_5 a$& 0.8114(10) & 0.5701(10) &  0.4214(17) &   &  0.3071(12)&\\
$\mu_6 a$& 0.9451(10) & 0.6658(09) &  0.4925(14) &   &  0.3588(10)&\\
$\mu_7 a$& 1.0647(09) & 0.7530(09) &  0.5579(14) &   &  0.4062(09)&\\\hline
$\mu_8 a$& 0.3892(14) & 0.2741(15) &  0.2038(40) &   &  0.1519(20)&\\
$\mu_9 a$& 0.5678(11) & 0.3984(12) &  0.2948(26) &   &  0.2160(16)&\\
\hline
\hline
\end{tabular}
\end{center}
\caption{\it Pseudoscalar meson masses $m_{\rm PS} a$ for all simulation points.}
\label{table:mpi}
\end{table}

\begin{table}[!t]
\begin{center}
\begin{tabular}{|c||c|c|c|c|c|c|}
\hline
\hline
$\beta$ & 5.70 &  5.85 & 6.00 & 6.10 & 6.20  & 6.45 \\   
\hline
\hline
&\multicolumn{6}{|c|}{$f_{\rm PS} a$ ($\kcpi$)}\\
\hline
$\mu_1 a$&0.0986(10)&0.0782(13)&0.0516(17)&0.0466(14)&0.0437(13)&0.0329(13)\\
$\mu_2 a$&0.1195(10)&0.0890(12)&0.0632(11)&0.0546(09)&0.0500(11)&0.0361(11)\\
$\mu_3 a$&0.1418(11)&0.1003(12)&0.0740(09)&0.0623(08)&0.0562(10)&\\
$\mu_4 a$&0.1685(11)&0.1149(12)&0.0859(09)&0.0716(08)&0.0637(10)&\\
$\mu_5 a$&0.1902(11)&0.1273(13)&0.0949(09)&0.0790(08)&0.0698(09)&\\
$\mu_6 a$&0.2112(12)&0.1390(14)&0.1029(10)&0.0858(08)&0.0754(09)&\\
$\mu_7 a$&0.2320(13)&0.1501(14)&0.1104(10)&0.0919(09)&0.0806(09)&\\
\hline                                                     
\hline                                                     
&\multicolumn{6}{|c|}{$f_{\rm PS} a$ ($\kcP$)}\\
\hline
$\mu_1 a$&0.1267(14) &0.0894(14) & 0.0689(27) & &0.0512(16) &\\
$\mu_2 a$&0.1345(13) &0.0947(13) & 0.0711(13) & &0.0532(13) &\\
$\mu_3 a$&0.1472(12) &0.1025(12) & 0.0763(10) & &0.0567(10) &\\
$\mu_4 a$&0.1697(12) &0.1159(12) & 0.0858(10) & &0.0633(08) &\\
$\mu_5 a$&0.1914(13) &0.1284(11) & 0.0944(10) & &0.0694(08) &\\
$\mu_6 a$&0.2134(14) &0.1402(11) & 0.1025(10) & &0.0751(08) &\\
$\mu_7 a$&0.2358(15) &0.1518(11) & 0.1100(10) & &0.0803(08) &\\\hline
$\mu_8 a$&0.1403(13) &0.0983(12) & 0.0734(11) & &0.0548(11) &\\
$\mu_9 a$&0.1589(12) &0.1095(12) & 0.0813(10) & &0.0601(09) &\\
\hline
\hline
\end{tabular}
\end{center}
\caption{\it Pseudoscalar meson decay constants $f_{\rm PS} a$ for all simulation points.}
\label{table:fpi}
\end{table}

The vector meson mass shows, in comparison to $f_{\rm PS}$, larger
statistical fluctuations at small masses. This fact, together with the
considerably larger lattice artifacts which affect the values obtained with 
the pion definition of $\kappa_c$, makes the continuum extrapolation
too difficult in this case. As a consequence we limit ourselves to
present only the data obtained with the PCAC definition of
$\kappa_c$. The vector meson mass has been extracted from the
correlators eqs.~(\ref{ca}) and (\ref{ct}) with local source and
Jacobi-smeared sink~\cite{Allton:1993wc}. We observe that the tensor correlator
systematically shows smaller statistical fluctuations and thus 
we report in table~\ref{table:mrho} only results obtained from this 
correlator. 
 
\begin{figure}[htb]
\vspace{-0.0cm}
\begin{center}
\epsfig{file=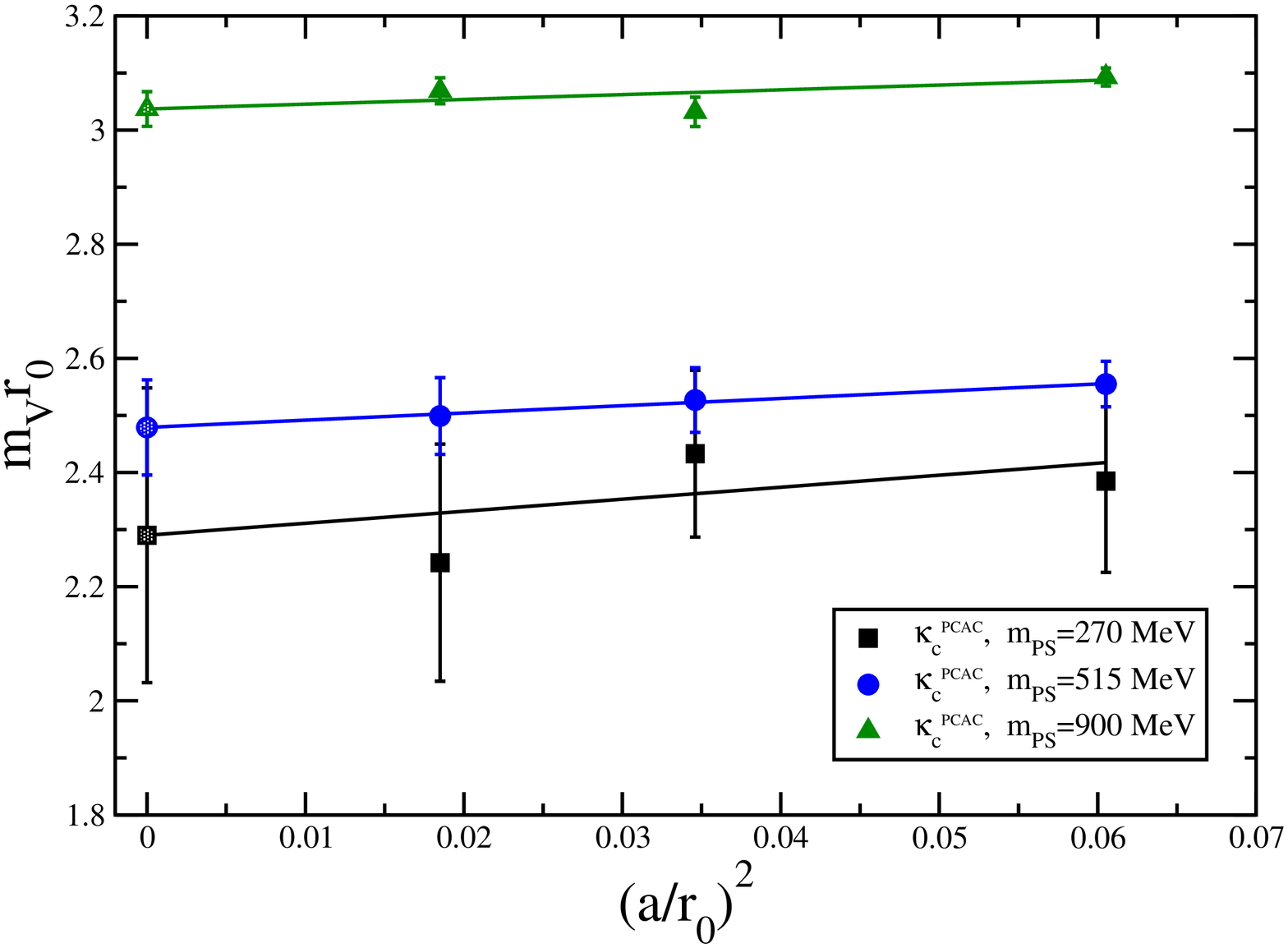,angle=270,width=0.8\linewidth}
\epsfig{file=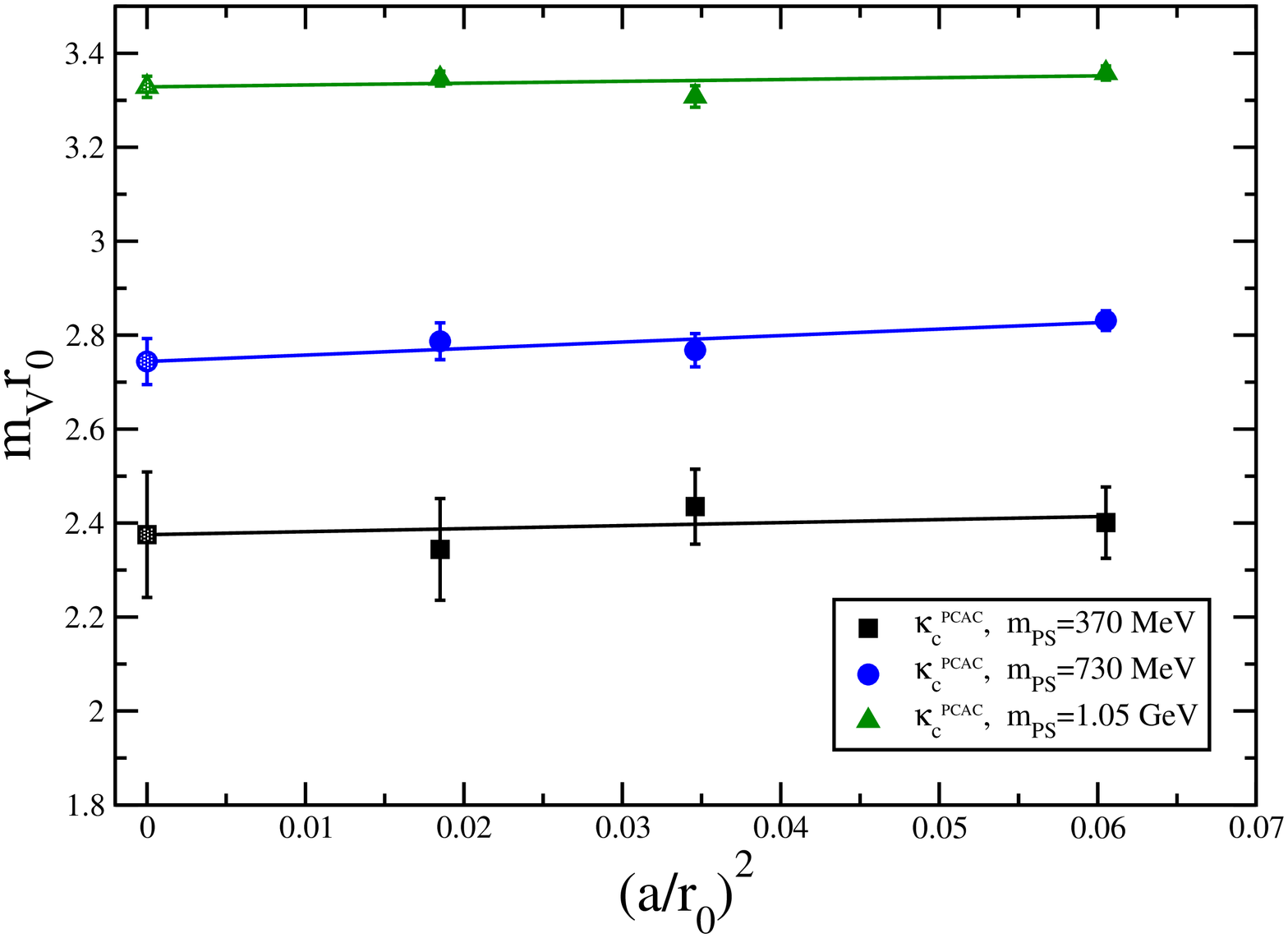,angle=270,width=0.8\linewidth}
\end{center}
\vspace{-0.0cm}
\caption{The vector meson mass as a function of $(a/r_0)^2$ as obtained from 
the PCAC definition of the critical mass. We also show a linear 
fit in $(a/r_0)^2$ of these quantity at fixed value of the charged 
pseudoscalar meson mass.
\label{fig:mv}}
\end{figure}

\begin{figure}[htb]
\vspace{-0.0cm}
\begin{center}
\epsfig{file=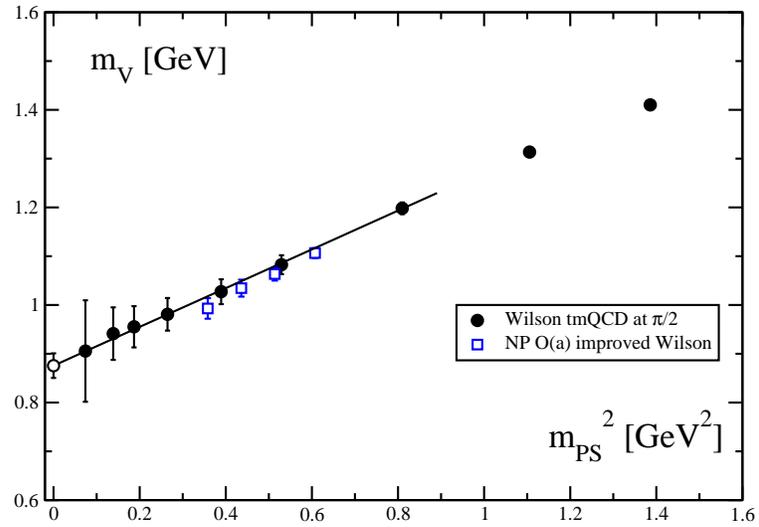,angle=270,width=0.8\linewidth}
\end{center}
\vspace{-0.0cm}
\caption{The continuum limit values of the vector meson mass (filled
  circles) as a function of the pseudoscalar meson mass squared. We also
  plot the continuum limit obtained with non-perturbatively improved
  Wilson fermions (open squares)~\cite{Garden:1999fg}.  
\label{fig:mvcont}}
\end{figure}

In fig.~\ref{fig:mv} we show the results for the vector meson mass as
function of $(a/r_0)^2$.
Again, even for small pseudoscalar meson masses, the behaviour of the 
vector meson mass is almost flat in $(a/r_0)^2$, indicating that $O(a^2)$
lattice artefacts  
are also small for this quantity. We perform linear fits of these data 
as function of $(a/r_0)^2$ which are represented by the lines in 
fig.~\ref{fig:mv}. The continuum extrapolated values for the vector meson
mass are presented in table~\ref{table:cont} and in 
fig.~\ref{fig:mvcont}. As a function of the 
pseudoscalar meson mass squared they show a linear behaviour without
signs of q$\chi$PT artefacts. In fig.~\ref{fig:mvcont} we also plot the
continuum values obtained with non-perturbatively improved Wilson 
fermions~\cite{Garden:1999fg}. 

In table~\ref{table:cont} the results of a linear extrapolation (performed with
the seven smallest masses) of our data to the chiral limit are given. Moreover,
we use this extrapolation to compute the values of $m_\rho$ and of $m_{K^*}$
(the latter in the $SU(3)$ symmetric limit). As already observed in quenched
calculations where the scale is determined through $r_0$, these values turn out
to be $10-15\%$ larger than the experimental values.  

\begin{table}[!t]
\begin{center}
\begin{tabular}{|c||c|c|c|c|}
\hline
\hline
$\beta$ & 5.70 &  5.85 & 6.00 &  6.20  \\   
\hline
\hline
&\multicolumn{4}{|c|}{$m_V a$ ($\kcP$)}\\
\hline
$\mu_1 a$&0.716(67)&0.589(43)& 0.458(30) &0.306(26)\\
$\mu_2 a$&0.773(36)&0.591(19)& 0.451(20) &0.317(20)\\
$\mu_3 a$&0.854(20)&0.628(09)& 0.467(12) &0.339(11)\\
$\mu_4 a$&0.973(15)&0.701(05)& 0.517(07) &0.378(05)\\
$\mu_5 a$&1.076(11)&0.765(04)& 0.560(06) &0.415(03)\\
$\mu_6 a$&1.178(09)&0.834(03)& 0.614(04) &0.452(02)\\
$\mu_7 a$&1.277(08)&0.902(03)& 0.666(03) &0.488(02)\\\hline
$\mu_8 a$&0.812(26)&0.606(13)& 0.464(14) &0.327(16)\\
$\mu_9 a$&0.919(14)&0.666(06)& 0.494(08) &0.359(07)\\
\hline
\hline
\end{tabular}
\end{center}
\caption{\it Vector meson masses $m_V a$ for all simulation points with
  the PCAC definition of $\kappa_c$.}
\label{table:mrho}
\end{table}

\begin{figure}[htb]
\begin{center}
\epsfig{file=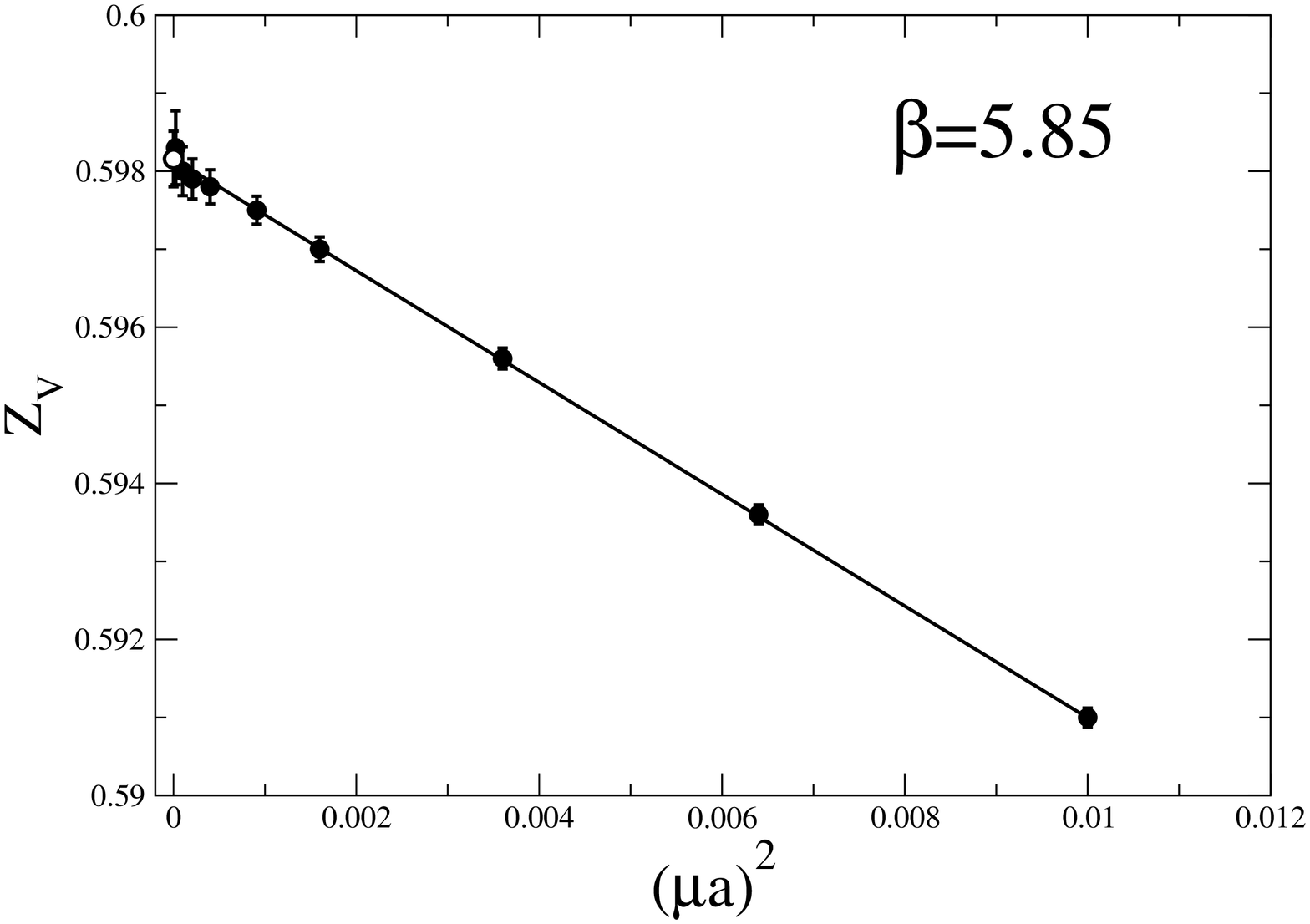,angle=270,width=0.8\linewidth}
~\\[-7.8cm]
\hspace*{4.1cm}
\epsfig{file=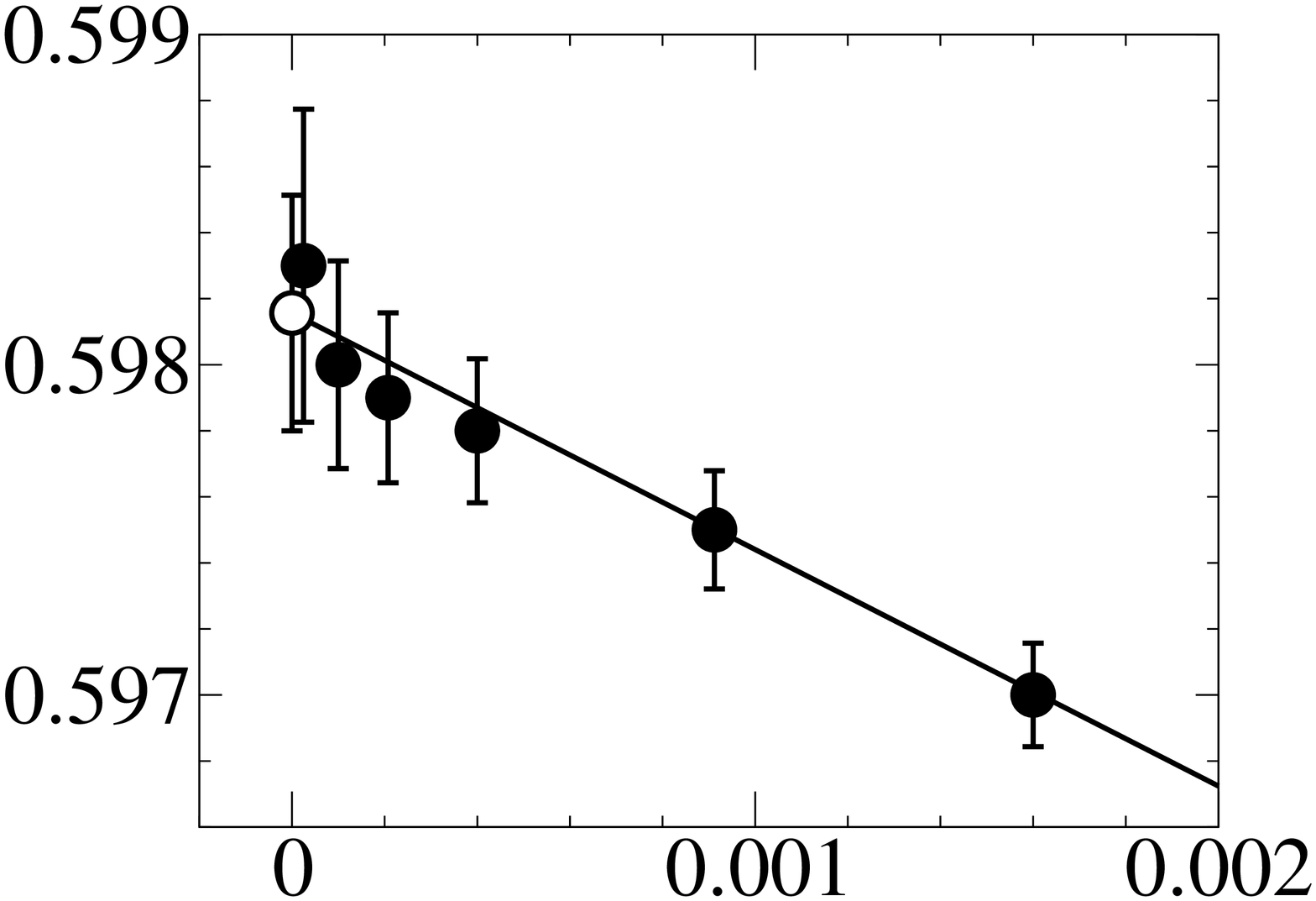,angle=270,width=0.31\linewidth}
\vspace*{4cm}
\end{center}
\caption{An example of the extrapolation of $Z_V$ to zero quark mass at 
$\beta=5.85$. The data show a linear behaviour in $(a\mu)^2$. 
\label{fig:zetav585}}
\end{figure}

As a last quantity, we computed the renormalisation constant $Z_V$ which
is defined in eq.~(\ref{eq:Z_V}) and is to be obtained in the chiral
limit. This renormalisation constant is expected to be equal,
apart from lattice artifacts, to the one obtained for pure Wilson fermions.
Also in this case we do not consider data obtained with the pion
definition of $\kappa_c$. The reason is, as was shown 
in~\cite{Bietenholz:2004wv}, that in this case the lattice 
artifacts affecting $Z_V$ become very large at small masses, thus 
preventing the possibility of performing a reliable extrapolation to the
chiral limit. 
Using the PCAC definition of $\kappa_c$, lattice artifacts are 
well under control also at small quark masses (with the exception of
$\beta=5.7$ where we prefer not to perform the chiral extrapolation).
In fig.~\ref{fig:zetav585} we provide one 
example of $Z_V$ as a function of the quark mass. It turns out that  
$Z_V$ is to a good approximation linear in $(a\mu)^2$. 
In table~\ref{table:zvscaling} we report the values of $Z_V$ extrapolated to 
the chiral limit together with the results from standard perturbation 
theory (SPT)\cite{Martinelli:1982mw} (where the parameter of the expansion
is $\alpha_s^{\rm SPT}=g_0^2/(4\pi)$, with $g_0^2=6/\beta$ the bare
lattice coupling constant) and from boosted perturbation
theory (BPT)\cite{Lepage:1992xa} (where the parameter of the expansion is 
$\alpha_s^{\rm BPT}=\alpha_s^{\rm SPT}/{\cal P}$, with 
${\cal P}=1/3\langle {\rm Re}\{\Tr P_{\mu\nu}\}\rangle$ 
the average value of the plaquette).  
Non-perturbative determinations with Wilson fermions using
different methods are present in the literature for $\beta=6.0$ and give
results in the range [0.57--0.74]. The spread is due to different $O(a)$
artifacts which affect the different determinations.

\begin{table}[!t]
\begin{center}
\begin{tabular}{|c||c|c|}
\hline
\hline
$m_{\rm PS}$ [GeV] & $f_{\rm PS}$ [GeV]& $m_V$ [GeV]\\
\hline
\hline
0.272 & 0.1500(66) & 0.904(102)\\
0.372 & 0.1538(45) & 0.937(53)\\
0.432 & 0.1572(40) & 0.955(42)\\
0.514 & 0.1631(36) & 0.978(33)\\
0.624 & 0.1724(33) & 1.027(26)\\
0.728 & 0.1823(31) & 1.083(19)\\
0.900 & 0.2002(29) & 1.198(12)\\
1.051 & 0.2161(28) & 1.313(09)\\
1.177 & 0.2283(28) & 1.410(07)\\
\hline
\hline
0.0   & 0.1439(37) & 0.876(25)\\
\hline
0.137 & 0.1452(37) & 0.883(25)\\
\hline
0.495 & 0.1617(45) & 0.973(27)\\
\hline
\hline
\end{tabular}
\end{center}
\caption{\it $f_{\rm PS}$ and $m_V$ in the continuum with the PCAC definition of
  $\kappa_c$. The values in the last three rows are obtained from a linear fit
  on the smallest 6 masses (7 in the case of $m_V$) and correspond to: 
  the values in the chiral limit (first row); $f_\pi$ and $m_\rho$ (second
  row); $f_K$ and $m_{K^*}$ in the $SU(3)$ symmetric limit (third row).}
\label{table:cont}
\end{table}

\begin{table}[!t]
\begin{center}
\begin{tabular}{|c|c|c|c|}
\hline
\hline
$\beta$ & $Z_V$ & $Z_V^{BPT}$ & $Z_V^{SPT}$\\
\hline
\hline
5.85& 0.5982(4) &0.69&0.82\\
6.0 & 0.6424(4) &0.71&0.83\\
6.2 & 0.6814(3) &0.73&0.83\\
\hline
\hline
\end{tabular}
\end{center}
\caption{\it $Z_V$ from the PCVC relation (PCAC definition of $\kappa_c$) 
  and comparison with standard perturbation theory (SPT) and
  boosted perturbation theory (BPT).}
\label{table:zvscaling}
\end{table}

\section{Conclusions}

In this paper, we have explored the potential of 
Wilson twisted mass fermions to reach small quark masses and fine 
values of the lattice spacing. Using the PCAC definition of the critical 
mass the scaling region is found to start already at $\beta=5.85$, 
for the observables investigated here and for masses down to $270$ MeV. 
In fact, with this definition of the critical mass, correlators, apart
from being automatically $O(a)$-improved, are only affected by $O(a^2)$
discretisation errors which remain small in the region of masses 
that satisfy the (order of magnitude) inequality 
$\mu > a^2 \Lambda_{QCD}^3$. In the case of the pion definition
the $O(a^2)$ effects are instead small only in the region $\mu > a
\Lambda_{QCD}^2$ and, as observed in~\cite{Bietenholz:2004wv} and in the
present work, may become quite relevant at masses in the range $250$-$450$
MeV and for the range of lattice spacings simulated here. In this
case the scaling region starts, for masses down to 270 MeV, 
around $\beta=6.0$. However, at the two smallest
quark masses, we included a point at $\beta=6.45$ in order to be sure
that we can safely control the continuum limit extrapolation. 

In the case of $f_{\rm PS}$ we have explicitly checked that 
both definitions of the critical mass lead {\it independently} to
consistent values in the continuum limit. 
For further simulations, the PCAC definition of $\kappa_c$ is
clearly preferable as it leads to considerably smaller lattice artifacts at
small quark masses, allowing for an enlargement of the scaling region.

The results of this paper clearly reveal that Wilson twisted mass fermions
allow for simulations at pseudoscalar meson masses of about $250$ MeV 
without running into problems with exceptional small eigenvalues or large
lattice artefacts. This statement holds, when the PCAC definition of
the critical mass is used. 
This is a very important lesson also for dynamical simulations, a lesson that
we could only learn through the detailed quenched study
performed here. In the dynamical case, however, small pseudoscalar meson 
masses are very difficult to simulate and further development of algorithms
is still crucial. For example, modern variants of the Hybrid Monte 
Carlo algorithm allow to simulate pseudoscalar meson masses of about $380$ MeV
\cite{Urbach:2005ji,Luscher:2004rx} without hitting the ``Berlin
Wall''~\cite{Jansen:2003nt,Bernard:2002pd}. We therefore believe 
that by using these algorithmic techniques  together with Wilson twisted mass 
fermions at maximal twist, realised by tuning the twist angle through
the PCAC definition of the critical mass, it becomes realistic to address the
small quark mass region in full, dynamical lattice QCD.

\section*{Acknowledgements}
We thank R.~Frezzotti, G.~C.~Rossi, L.~Scorzato and U.~Wenger
for many useful discussions. 
The computer centres at NIC/DESY Zeuthen, NIC at Forschungszentrum
J{\"u}lich and HLRN provided the necessary technical help and computer
resources. This work was supported by the DFG 
Sonderforschungsbereich/Transregio SFB/TR9-03.

\bibliographystyle{JHEP}
\bibliography{scaling}

\providecommand{\href}[2]{#2}\begingroup\raggedright\begin{thebibliography}{10}

\bibitem{Frezzotti:2000nk}
{\bf ALPHA} Collaboration, R.~Frezzotti, P.~A. Grassi, S.~Sint, and P.~Weisz,
  {\it Lattice {QCD} with a chirally twisted mass term},  {\em JHEP} {\bf 08}
  (2001) 058, [\href{http://xxx.lanl.gov/abs/hep-lat/0101001}{{\tt
  hep-lat/0101001}}].

\bibitem{Frezzotti:2003ni}
R.~Frezzotti and G.~C. Rossi, {\it Chirally improving {Wilson} fermions. {I}:
  {O(a)} improvement},  {\em JHEP} {\bf 08} (2004) 007,
  [\href{http://xxx.lanl.gov/abs/hep-lat/0306014}{{\tt hep-lat/0306014}}].

\bibitem{Jansen:2003ir}
{\bf \xlf} Collaboration, K.~Jansen, A.~Shindler, C.~Urbach, and I.~Wetzorke,
  {\it Scaling test for {Wilson} twisted mass {QCD}},  {\em Phys. Lett.} {\bf
  B586} (2004) 432--438, [\href{http://xxx.lanl.gov/abs/hep-lat/0312013}{{\tt
  hep-lat/0312013}}].

\bibitem{Bietenholz:2004wv}
{\bf \xlf} Collaboration, W.~Bietenholz {\em et~al.}, {\it Going chiral:
  Overlap versus twisted mass fermions},  {\em JHEP} {\bf 12} (2004) 044,
  [\href{http://xxx.lanl.gov/abs/hep-lat/0411001}{{\tt hep-lat/0411001}}].

\bibitem{Jansen:2005gf}
{\bf \xlf} Collaboration, K.~Jansen, M.~Papinutto, A.~Shindler, C.~Urbach, and
  I.~Wetzorke, {\it Light quarks with twisted mass fermions},  {\em Phys.
  Lett.} {\bf B619} (2005) 184--191,
  [\href{http://xxx.lanl.gov/abs/hep-lat/0503031}{{\tt hep-lat/0503031}}].

\bibitem{Abdel-Rehim:2004gx}
A.~M. Abdel-Rehim and R.~Lewis, {\it Twisted mass {QCD} for the pion
  electromagnetic form factor},  {\em Phys. Rev.} {\bf D71} (2005) 014503,
  [\href{http://xxx.lanl.gov/abs/hep-lat/0410047}{{\tt hep-lat/0410047}}].

\bibitem{Abdel-Rehim:2005gz}
A.~M. Abdel-Rehim, R.~Lewis, and R.~M. Woloshyn, {\it Spectrum of quenched
  twisted mass lattice {QCD} at maximal twist},  {\em Phys. Rev.} {\bf D71}
  (2005) 094505, [\href{http://xxx.lanl.gov/abs/hep-lat/0503007}{{\tt
  hep-lat/0503007}}].

\bibitem{Farchioni:2004us}
F.~Farchioni {\em et~al.}, {\it Twisted mass quarks and the phase structure of
  lattice {QCD}},  {\em Eur. Phys. J.} {\bf C39} (2005) 421--433,
  [\href{http://xxx.lanl.gov/abs/hep-lat/0406039}{{\tt hep-lat/0406039}}].

\bibitem{Farchioni:2004ma}
F.~Farchioni {\em et~al.}, {\it Exploring the phase structure of lattice
  {{QCD}} with twisted mass quarks},  {\em Nucl. Phys. Proc. Suppl.} {\bf 140}
  (2005) 240--245, [\href{http://xxx.lanl.gov/abs/hep-lat/0409098}{{\tt
  hep-lat/0409098}}].

\bibitem{Farchioni:2004fs}
F.~Farchioni {\em et~al.}, {\it The phase structure of lattice {QCD} with
  {Wilson} quarks and renormalization group improved gluons},  {\em {\rm
  accepted for publication in} Eur. Phys. J.} {\bf C} (2005)
  [\href{http://xxx.lanl.gov/abs/hep-lat/0410031}{{\tt hep-lat/0410031}}].

\bibitem{Ilgenfritz:2003gw}
E.-M. Ilgenfritz, W.~Kerler, M.~M{\"u}ller-Preu{\ss}ker, A.~Sternbeck, and
  H.~St{\"u}ben, {\it A numerical reinvestigation of the {Aoki} phase with
  {N(f)} = 2 {Wilson} fermions at zero temperature},  {\em Phys. Rev.} {\bf
  D69} (2004) 074511, [\href{http://xxx.lanl.gov/abs/hep-lat/0309057}{{\tt
  hep-lat/0309057}}].

\bibitem{Sternbeck:2003gy}
A.~Sternbeck, E.-M. Ilgenfritz, W.~Kerler, M.~M{\"u}ller-Preu{\ss}ker, and
  H.~St{\"u}ben, {\it The {Aoki} phase for {N(f)} = 2 {Wilson} fermions
  revisited},  {\em Nucl. Phys. Proc. Suppl.} {\bf 129} (2004) 898--900,
  [\href{http://xxx.lanl.gov/abs/hep-lat/0309059}{{\tt hep-lat/0309059}}].

\bibitem{Bietenholz:2004sa}
{\bf \xlf} Collaboration, W.~Bietenholz {\em et~al.}, {\it Comparison between
  overlap and twisted mass fermions towards the chiral limit},  {\em Nucl.
  Phys. Proc. Suppl.} {\bf 140} (2005) 683--685,
  [\href{http://xxx.lanl.gov/abs/hep-lat/0409109}{{\tt hep-lat/0409109}}].

\bibitem{Aoki:2004ta}
S.~Aoki and O.~Bar, {\it Twisted-mass {QCD}, {O(a)} improvement and {Wilson}
  chiral perturbation theory},  {\em Phys. Rev.} {\bf D70} (2004) 116011,
  [\href{http://xxx.lanl.gov/abs/hep-lat/0409006}{{\tt hep-lat/0409006}}].

\bibitem{Sharpe:2004ny}
S.~R. Sharpe and J.~M.~S. Wu, {\it Twisted mass chiral perturbation theory at
  next-to-leading order},  {\em Phys. Rev.} {\bf D71} (2005) 074501,
  [\href{http://xxx.lanl.gov/abs/hep-lat/0411021}{{\tt hep-lat/0411021}}].

\bibitem{Frezzotti:2005gi}
R.~Frezzotti, G.~Martinelli, M.~Papinutto, and G.~C. Rossi, {\it Reducing
  cutoff effects in maximally twisted lattice {QCD} close to the chiral limit},
   \href{http://xxx.lanl.gov/abs/hep-lat/0503034}{{\tt hep-lat/0503034}}.

\bibitem{Frezzotti:2001du}
R.~Frezzotti and S.~Sint, {\it Some remarks on {O(a)} improved twisted mass
  {QCD}},  {\em Nucl. Phys. Proc. Suppl.} {\bf 106} (2002) 814--816,
  [\href{http://xxx.lanl.gov/abs/hep-lat/0110140}{{\tt hep-lat/0110140}}].

\bibitem{DellaMorte:2001tu}
M.~Della~Morte, R.~Frezzotti, and J.~Heitger, {\it Quenched twisted mass {QCD}
  at small quark masses and in large volume},  {\em Nucl. Phys. Proc. Suppl.}
  {\bf 106} (2002) 260--262,
  [\href{http://xxx.lanl.gov/abs/hep-lat/0110166}{{\tt hep-lat/0110166}}].

\bibitem{Garden:1999fg}
{\bf ALPHA} Collaboration, J.~Garden, J.~Heitger, R.~Sommer, and H.~Wittig,
  {\it Precision computation of the strange quark's mass in quenched {QCD}},
  {\em Nucl. Phys.} {\bf B571} (2000) 237--256,
  [\href{http://xxx.lanl.gov/abs/hep-lat/9906013}{{\tt hep-lat/9906013}}].

\bibitem{Heitger:2000ay}
{\bf ALPHA} Collaboration, J.~Heitger, R.~Sommer, and H.~Wittig, {\it Effective
  chiral lagrangians and lattice {{QCD}}},  {\em Nucl. Phys.} {\bf B588} (2000)
  377--399, [\href{http://xxx.lanl.gov/abs/hep-lat/0006026}{{\tt
  hep-lat/0006026}}]. and references therein.

\bibitem{Allton:1993wc}
{\bf UK{QCD}} Collaboration, C.~R. Allton {\em et~al.}, {\it Gauge invariant
  smearing and matrix correlators using {Wilson} fermions at beta = 6.2},  {\em
  Phys. Rev.} {\bf D47} (1993) 5128--5137,
  [\href{http://xxx.lanl.gov/abs/hep-lat/9303009}{{\tt hep-lat/9303009}}].

\bibitem{Martinelli:1982mw}
G.~Martinelli and Y.-C. Zhang, {\it The connection between local operators on
  the lattice and in the continuum and its relation to meson decay constants},
  {\em Phys. Lett.} {\bf B123} (1983) 433.

\bibitem{Lepage:1992xa}
G.~P. Lepage and P.~B. Mackenzie, {\it On the viability of lattice perturbation
  theory},  {\em Phys. Rev.} {\bf D48} (1993) 2250--2264,
  [\href{http://xxx.lanl.gov/abs/hep-lat/9209022}{{\tt hep-lat/9209022}}].

\bibitem{Urbach:2005ji}
C.~Urbach, K.~Jansen, A.~Shindler, and U.~Wenger, {\it {HMC} algorithm with
  multiple time scale integration and mass preconditioning},
  \href{http://xxx.lanl.gov/abs/hep-lat/0506011}{{\tt hep-lat/0506011}}.

\bibitem{Luscher:2004rx}
M.~L{\"u}scher, {\it Schwarz-preconditioned {HMC} algorithm for two-flavour
  lattice {QCD}},  {\em Comput. Phys. Commun.} {\bf 165} (2005) 199,
  [\href{http://xxx.lanl.gov/abs/hep-lat/0409106}{{\tt hep-lat/0409106}}].

\bibitem{Jansen:2003nt}
K.~Jansen, {\it Actions for dynamical fermion simulations: Are we ready to
  go?},  {\em Nucl. Phys. Proc. Suppl.} {\bf 129} (2004) 3--16,
  [\href{http://xxx.lanl.gov/abs/hep-lat/0311039}{{\tt hep-lat/0311039}}].

\bibitem{Bernard:2002pd}
C.~Bernard {\em et~al.}, {\it Panel discussion on the cost of dynamical quark
  simulations},  {\em Nucl. Phys. Proc. Suppl.} {\bf 106} (2002) 199--205.

\bibitem{Freund}
R.~Freund {\em in Numerical Linear Algebra, L.\ Reichel, A.\ Ruttan and R.S.\
  Varga (eds.)} (1993) p. 101.

\bibitem{Glassner:1996gz}
U.~Gl{\"a}ssner {\em et~al.}, {\it How to compute {G}reen's functions for
  entire mass trajectories within {K}rylov solvers},
  \href{http://xxx.lanl.gov/abs/hep-lat/9605008}{{\tt hep-lat/9605008}}.

\bibitem{Jegerlehner:1997rn}
B.~Jegerlehner, {\it Multiple mass solvers},  {\em Nucl. Phys. Proc. Suppl.}
  {\bf 63} (1998) 958--960, [\href{http://xxx.lanl.gov/abs/hep-lat/9708029,
  hep-lat/9612014}{{\tt hep-lat/9708029, hep-lat/9612014}}].

\end{thebibliography}\endgroup

\newpage
\begin{appendix}
\section*{Appendix}
\subsection*{Multiple mass solver for twisted mass fermions}
In this appendix we show that within the Wilson twisted mass fermion
formulation it is possible to apply the multiple mass solver (MMS)
\cite{Freund,Glassner:1996gz,Jegerlehner:1997rn} method to the 
conjugate gradient (CG) algorithm. We will call this algorithm CG-M
and give here the details of the implementation. 

The advantage of the MMS is that it allows the computation of the solution of 
the following linear system
\be
(A+\sigma)\;x-b=0
\ee
for several values of $\sigma$ simultaneously, using only as many matrix-vector
operations as the solution of a single value of $\sigma$ requires.

We want to invert the Wilson twisted mass operator at a certain twisted mass
$\mu_0$ obtaining automatically all the solutions for other 
twisted masses $\mu_k$ (with $|\mu_k| \ge |\mu_0|$).
In the so called twisted basis the Wilson twisted mass operator is
\be
D_{\rm tm} = D + i\mu_k\gamma_5\tau^3, \qquad k=1,\ldots,N_m
\ee
where $D$ is the standard massive Wilson operator $D = D_W + m_0$ (see
eq.~(\ref{tmaction}) and (\ref{Dw}))
and $N_m$ is the number of additional twisted masses.
The operator  can be splitted as
\be
D_{\rm tm} = D_{\rm tm}^{(0)} + i(\mu_k-\mu_0)\gamma_5\tau^3, 
\qquad D_{\rm tm}^{(0)}  = D + i\mu_0\gamma_5\tau^3 \;.
\ee
The trivial observation is that 
\be
D_{\rm tm}D_{\rm tm}^{\dagger} = D_{\rm tm}^{(0)}D_{\rm tm}^{(0)\dagger} 
+ \mu_k^2 - \mu_0^2\;,
\ee
where we have used $\gamma_5D_W\gamma_5 = D_W^{\dagger}$.
Now clearly we have a shifted linear system $(A+\sigma_k)x-b=0$ 
with $A = D_{\rm tm}^{(0)}D_{\rm tm}^{(0)\dagger}$ and 
$\sigma_k = \mu_k^2 - \mu_0^2$.
We describe now the CG-M algorithm in order to solve the problem
$(A+\sigma_k)x-b=0$.
The lower index indicates the iteration steps of the solver, while the upper
index $k$ refers to the shifted problem with $\sigma_k$. 

\begin{align*}
&{\rm CG-M~~Algorithm} \\[2mm]
& x_0^k = 0, r_0 = p_0^k = b,
\alpha_{-1} = \zeta_{-1}^k = \zeta_0^k = 1, \beta_0^k = \beta_0 = 0 \\
& {\rm for}~~i=0,1,2,\cdots \\
& \quad \alpha_n = {(r_n, r_n)\over (p_n, A p_n)} \\
& \quad \zeta_{n+1}^k = {\zeta^k_n  \alpha_{n-1} \over 
\alpha_n \beta_n(1 - \frac{\zeta_n^k}{\zeta^k_{n-1}}) + \alpha_{n-1}
(1-\sigma_k\alpha_n)} \\
& \quad \alpha^k_n = \alpha_n {\zeta_{n+1}^k \over \zeta_n^k} \\
& \quad x_{n+1}^k = x_n^k + \alpha_n^k p_n^k \\
& \quad x_{n+1} = x_n + \alpha_n p_n \\
& \quad r_{n+1} = r_n - \alpha_n Ap_n \\
& \quad {\rm convergence~ check} \\
& \quad \beta_{n+1} = {(r_{n+1}, r_{n+1})\over (r_n, r_n)} \\
& \quad p_{n+1} = r_{n+1} + \beta_{n+1} p_n \\
& \quad \beta_{n+1}^k = \beta_{n+1} {\zeta_{n+1}^k \alpha_n^k \over \zeta_n^k 
\alpha_n} \\
& \quad p_{n+1}^k = \zeta_{n+1}^k r_{n+1} + \beta_{n+1}^k p_n^k\\
& {\rm end~for}
\end{align*}
We give here the algorithm explicitly again, since it has a different
definition of $\zeta_{n+1}^k$ compared to the one of \cite{Jegerlehner:1997rn}.
This version allows to avoid roundoff errors when 
$\sigma_k = \mu_k^2 - \mu_0^2$ becomes too large. 

We remind that when using a MMS the eventual preconditioning has to retain the
shifted structure of the linear system. This means for example that it is not
compatible with even-odd preconditioning.
\end{appendix}

\end{document}